\documentclass[aps,prl,reprint,groupedaddress,longbibliography]{revtex4-1}
\usepackage[pdftex]{graphicx}
\usepackage{graphicx}
\usepackage[utf8]{inputenc}
\usepackage{amsmath}
\usepackage{braket}

\newcommand{\erf}{\mathrm{erf}}

\begin{document}

\title{Measuring adiabaticity in non-equilibrium quantum systems}

\author{A. H. Skelt, R. W. Godby and I. D'Amico}
\affiliation{Department of Physics, University of York, York YO10 5DD, United Kingdom}

\date{\today}

\begin{abstract}
Understanding out-of-equilibrium quantum dynamics is a critical outstanding problem, with key questions regarding characterizing adiabaticity for applications in quantum technologies.  We show how the metric-space approach to quantum mechanics naturally characterizes regimes of quantum dynamics, and provides an appealingly visual tool for assessing their degree of adiabaticity. Further, the dynamic trajectories of quantum systems in metric space suggest a lack of ``ergodicity'', thus providing a better understanding of the fundamental one-to-one mapping between densities and wavefunctions.
\end{abstract}

\pacs{}

\maketitle
\section{Introduction}
Characterization and control of the dynamics of quantum systems is essential for the development of quantum technologies such as quantum computation and simulation, and for the emerging field of quantum thermodynamics \cite{Kosloff2013,Vinjanampathy2016,Millen2016}.  This has triggered increasing interest in understanding the properties of quantum systems out-of-equilibrium, and in identifying signatures of adiabatic behavior.  Many applications in quantum technologies require adiabatic processes.  These range from theoretical concepts such as Landau-Zener transitions \cite{Shevchenko2010}, Berry phase accumulation \cite{Berry1984} and the quantum Hall effect \cite{Laughlin1981}, to experimental techniques of adiabatic passage protocols \cite{Bergmann1998,Vitanov2001,Greentree2004}, for example.  In fact adiabatic dynamics may be used to efficiently perform a desired quantum evolution, as in adiabatic quantum computation \cite{Fahri2001,Roland2002,Das2002,Albash2016}, or to avoid quantum friction in the production of quantum work \cite{Plastina2014}. Adiabatic quantum dynamics through quantum annealing is indeed the motor of the commercial D-Wave  `quantum' computer \cite{Johnson2011,Gibney2017}. Knowledge of the degree of adiabaticity in non-equilibrium dynamics is also important for time-dependent (TD) density functional theory (DFT), as effective density functionals are, at present, available only in the (near) adiabatic regime.

The quantum adiabatic theorem \cite{BornFock}, which states that for a Hamiltonian varying slowly enough a system initially in equilibrium will remain in its instantaneous ground state, properly characterizes the dynamics of quantum systems.  However the commonly used quantum adiabatic criterion (QAC) \cite{Tong2005,Marzlin2004} is not always accurate in characterizing the degree of adiabaticity, with recent discussions and experiments showing the criterion is not always sufficient or necessary \cite{Marzlin2004,Ortigoso2012,Comparat2009,Li2016,Du2008,Jansen2007,Amin2009}. The QAC is based on perturbation theory and usually only considers two eigenstates, which adds to its limitations.

In this paper, we demonstrate the use of `natural' metrics~\cite{DAmico, S&D} as an efficient yet simple tool for characterizing the degree of adiabaticity in quantum systems.  The metrics, which avoid several limitations of the QAC, are applied to a broad range of systems and provide insight into the degree of adiabaticity, even when the outcome from the QAC is questionable.
Further, a better understanding of the one-to-one mapping between TD densities and wavefunctions, core to TDDFT \cite{Runge}, is particularly needed.  The wavefunction-density relationship, is a mapping between metric spaces~\cite{DAmico}, so the metric-based analysis gives us a fitting tool with which to study it~\cite{DAmico, S&D, S&D2}.

We will look at the relationship in metric space between densities and wavefunctions for a diverse set of one-dimensional systems, and study how this relationship changes as the systems become time-dependent and evolve out of equilibrium. We address the fundamental questions: How are ground states characterized by metric spaces? Is the quantum dynamics of systems ``ergodic'' within a metric space? What is the signature of out-of-equilibrium regimes in metric spaces? Can metric spaces efficiently characterize different dynamic regimes, and in particular the crossover between adiabaticity and non-adiabaticity?

In Refs.~\cite{DAmico} and \cite{S&D}, the concept of `natural' metrics -- directly arising from conservation laws -- was introduced. The `natural' metrics which measure the distance between two $N$-particle wavefunctions (normalized to $N$), or two $N$-particle densities are, respectively~\cite{DAmico},
\begin{equation}
  \label{wf full metric}
  D_{\psi}\left(\psi_1, \psi_2 \right) = \left[ 2N - 2 \left| \smallint \psi^*_1 \psi_2 \, dr_1 \ldots dr_N \right| \right]^{\frac{1}{2}} \, ;
\end{equation}
\begin{equation}
\label{density metric}
  D_{n}\left( n_1, n_2 \right) = \smallint |n_1(\textbf{r}) - n_2(\textbf{r})|d^3\textbf{r} \, .
\end{equation}

\section{Ground state systems}
\begin{figure}
\includegraphics[width=0.48\textwidth]{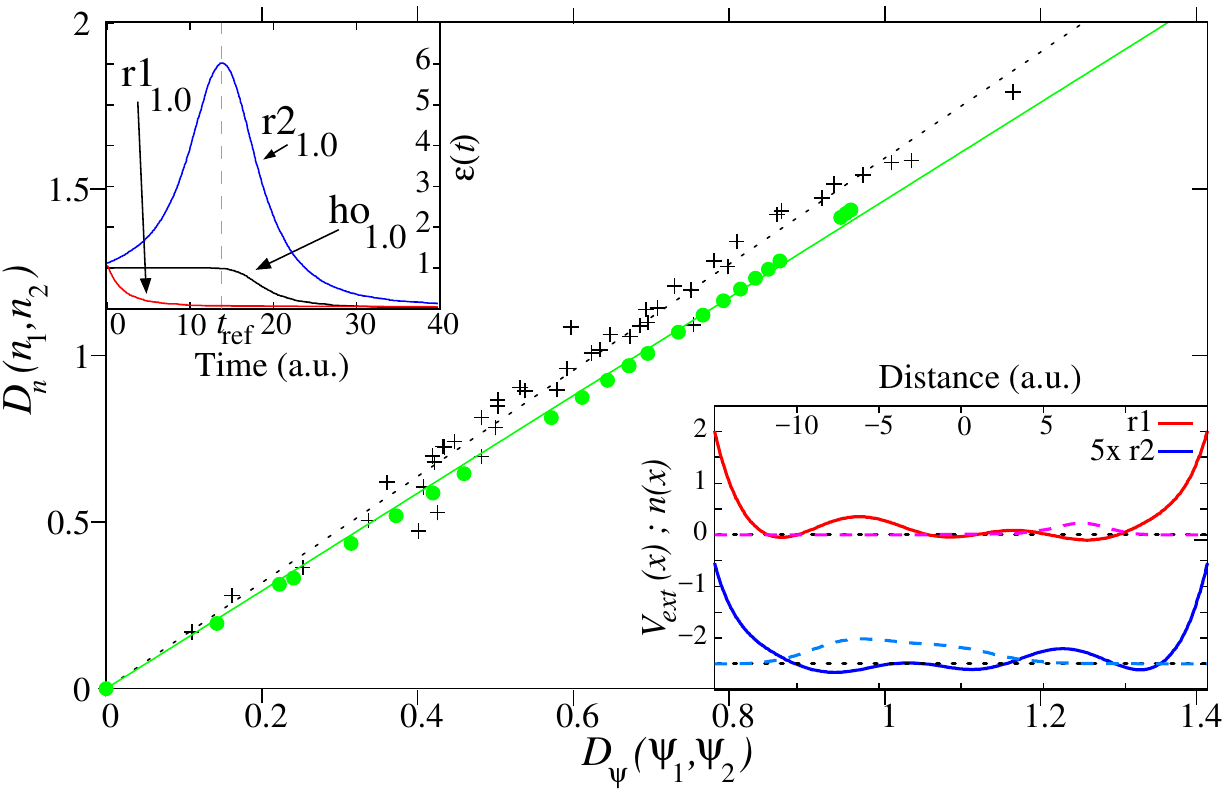}
  \caption{\textit{Main panel:} Metrics $D_{n} $ vs $D_{\psi}$  for  10 random single electron systems (black crosses) and 23 simple harmonic oscillator systems (green circles) in their GSs. $D_{n}/D_{\psi}$ is approximately linear with similar gradients of 1.59 and 1.43 respectively.
\textit{Inset (lower):} Two examples of our random potentials (solid lines) and their GS densities (dashed lines).  These are used for the TD study: system r1 (red, $\Lambda = 0.5$) and system r2 (blue, $\Lambda = 0.1$, the spatial reflection of r1 divided by five); the curves are displaced vertically so that the GS energies lie at 0 and $-2.5$ on the vertical axis, respectively.
\textit{Inset (upper):} TD adiabaticity parameter $\epsilon(t)$ (Eq.~\ref{epsilon}) for the three time dependent systems (r1, r2 and a harmonic oscillator, ho) corresponding to $\epsilon(0) = 1.0$.   The vertical gray dashed line shows the reference time, $t_{\mathrm{ref}}$, used in Fig.~\ref{fig:den_plot}.}
\label{fig:gs_SHO_rand}
\end{figure}

We explore the mapping between ground state (GS) particle densities and the corresponding wavefunctions for single-particle systems, beginning with harmonic oscillators and then moving onto more complex, randomly generated systems.
By inserting the analytic GSs of two harmonic oscillators into Eqs. (\ref{wf full metric}) and (\ref{density metric}), the ratio of the metrics may be written exactly as
\begin{equation}
    \label{SHOgrad}
    \frac{D_{n}\left( n_1, n_2 \right)}{D_{\psi}\left(\psi_1, \psi_2 \right)} = \frac{2 \left[ \erf \left( \sqrt{\frac{\nu\ln(\nu)}{2(\nu-1)}} \right) - \erf \left( \sqrt{\frac{\ln(\nu)}{2(\nu-1)}} \right) \right] } {\sqrt{2 - \frac{2^{3/2}\nu^{1/4}}{(\nu+1)^{1/2}}}},
\end{equation}
where $\nu=\omega_1/\omega_2$ is the ratio of the frequencies of the two oscillators.
Expanding this about $\nu=1$, we obtain $D_{n} \left( n_1, n_2 \right)/D_{\psi}\left(\psi_1, \psi_2 \right) = 4/\sqrt{e\pi} + O(v-1)^2$
where $e$ is the base of natural logarithms, demonstrating a linear relationship with gradient $4/\sqrt{e\pi} \approx 1.37$ when $\omega_1 \approx \omega_2$. Our numerical results confirm the linear relationship even for $|\nu|\gg 1$: we compare  23 simple harmonic oscillators with a range of frequencies ($\omega$) from 0.05 to 2.20 a.u. \footnote{We use atomic units, $\hbar=m=1$} with a reference oscillator for which $\omega = 0.1$.  This yields the green circles in Fig.~\ref{fig:gs_SHO_rand} (main panel) which are well described by a straight line with gradient 1.43.

Next, we consider systems  with smooth, random, confining potentials. These are generated using a Fourier series with random coefficients, together with an $x^{10}$ potential to gently confine the electrons overall: $V_{ext}(x) =  {x^{10}}/{10^{11}} + \Lambda \sum_{n=1}^3 \left( a_n \cos \tfrac{n \pi x}{L} + b_n \sin \tfrac{n \pi x}{L} \right)$.
Here $L$ is half the system size, and the $a_n$ and $b_n$ are drawn from a uniform distribution between $-\frac{L}{3}$ and $\frac{L}{3}$.  The scaling factor $\Lambda$ is used to adjust the confining strength of the potential microwells, allowing different regimes of electron localization to be explored.  By using the Fourier series, we generate a wide range of potentials that vary in multiple parameters, unlike the Hamiltonians in Ref. \cite{DAmico} which vary in only one parameter when comparing systems.

Fig.~\ref{fig:gs_SHO_rand} (lower inset) shows examples of two random potentials. For the GS study we used a family of ten random potentials with $\Lambda = 0.1$ and $L=15$ a.u. We solve the Schrödinger equation for our systems using the SPiDEA code \footnote{\label{SPiDEA}J. Wetherell, unpublished; subsequently incorporated into the iDEA code suite \cite{Hodgson}} to obtain the exact GS wavefunctions and densities, from which $D_{\psi}$ and $D_{n} $ are calculated using Eqs.~\ref{wf full metric} and \ref{density metric}. Fig.~\ref{fig:gs_SHO_rand} (main panel) shows $D_{n} $ against $D_{\psi}$ for all 45 pairs of systems in the family (black crosses).  The points lie close to a straight line through the origin with gradient 1.59, deviating slightly solely to reach the combination of the maximum values of $D_{n} $ and $D_{\psi}$ (2 and $\sqrt{2}$ respectively, top right-hand corner of the graph).

Ref.~\cite{DAmico} found a similar quasi-linear relationship between $D_{n}$ and $D_{\psi}$ for three families of systems, with the gradient depending on the number of particles, $N$. There, the families of systems were each generated by varying a single parameter in the Hamiltonian (e.g. the confining frequency for Hooke's atoms), while here a diverse range of systems are explored for $N=1$ \footnote{Preliminary results for random potentials with $N$=2 also show a quasi-linear relationship.}.

\section{Time dependent systems}
The quasi-linear relationship of $D_{n} \left( n_1, n_2 \right)$ and $D_{\psi}\left(\psi_1, \psi_2 \right)$, for GSs, may therefore become a tool to identify whether the time-dependence of a quantum system is adiabatic \footnote{For the evolution of GSs this quasi-linear relationship indicates both equilibrium and adiabaticity; Ref.~\cite{S&D2} suggests that a similar relationship may hold also for excited states, hence the proposed method could be extended to any eigenstate}.  For $N=1$ we take this relationship to be 1.5 as an average of the harmonic systems and random systems. The quantum adiabatic theorem \cite{BornFock} states that for a Hamiltonian varying slowly enough, a system initially at equilibrium will remain in an eigenstate of the instantaneous Hamiltonian, $\hat{H}$.  Quantification of the adiabatic theorem is traditionally based on the criterion \cite{Marzlin2004,Tong2005}
\begin{equation}
\label{epsilon}
\epsilon(t) = \frac{\left| \bra{m} \dot{H} \ket{n} \right|}{\left( \left| E_n - E_m \right| \right)^2} \ll 1
\end{equation}
where $n$ is the perfectly adiabatically-evolving original eigenstate, $m$ corresponds to another eigenstate of the instantaneous Hamiltonian, and typically  $m=n \pm 1$. In recent years, debate has opened up about the validity and sufficiency of the quantum adiabatic criterion, with some conclusions showing it to break down for specially crafted systems with oscillating terms in the Hamiltonian \cite{Marzlin2004,Ortigoso2012,Comparat2009}.  However the question remains open \cite{Li2016}. Furthermore this criterion is derived from perturbation theory which may not be applicable for stronger perturbations.
Here we propose metrics to provide a graphical method of determining adiabaticity which avoids the limitations of $\epsilon(t)$. Metrics are non-perturbative and automatically consider all eigenstates, providing further insight into the dynamics of the system not available from $\epsilon(t)$ \footnote{Euclidean distances between wavefunctions have previously been used \cite{Li2016} to study the validity of $\epsilon(t)$, but are inappropriately sensitive to a physically-irrelevant overall phase-change of the state; the metrics used here are tailored to avoid this shortcoming \cite{DAmico}}.

To explore adiabaticity, we use the SPiDEA code to  turn on a uniform electric field increasing linearly with time with a rate $p$, making the Hamiltonian of our systems $\hat{H}(x,t) = -\frac{1}{2}\frac{\partial^2}{\partial x^2} + V_{ext}(x) - ptx$.

We evaluate the distances between a system's initial GS, $\psi(0)$, instantaneous GS, $\psi_{GS}(t)$, and time dependent state, $\psi(t)$; we obtain $D_\psi(\psi(0),\psi(t))$, $D_\psi(\psi(0),\psi_{GS}(t))$ and $D_\psi(\psi_{GS}(t),\psi(t))$ from Eq.~\ref{wf full metric}, and corresponding expressions for the density from Eq.~\ref{density metric}.

We focus on three initial systems, r1, r2 (from Fig. \ref{fig:gs_SHO_rand}), and a harmonic oscillator with $\omega = 0.2$ (ho).  Each system is perturbed at two different rates. The six systems span a rich spectrum of behaviors, showing the transition from the harmonic system, ho, through the random potential r1, with a harmonic-like microwell which also allows for mild tunneling into the neighboring well, to the random potential r2, with a GS delocalized over multiple microwells \cite{Skelt2017-2}.
We choose the perturbation rates $p$ so that the initial adiabaticity parameter $\epsilon(0)$  (from Eq.~\ref{epsilon}) takes the same two values for all three initial potentials \footnote{For $\epsilon(0)=0.01$, the values of $p$ are 2.530, 0.15 and 0.025 for the ho, r1 and r2 systems, respectively, while for $\epsilon(0)=1.0$ the values of $p$ are 100 times greater.}.

By definition, if an adiabatic regime is reached, our systems should remain in the GS of the instantaneous Hamiltonian at every time step: from the findings in Fig.~\ref{fig:gs_SHO_rand}, we then expect the dynamics in metric space of such systems to be described by a linear relationship between $D_n(n(0),n(t))$ and $D_\psi(\psi(0),\psi(t))$. By using three types of graphs, we will study how such a regime is entered/exited and, in general, characterized in metric space. These graphs deliver complementary perspectives on the systems' time evolution and adiabaticity.

\begin{figure}
\includegraphics[width=0.48\textwidth]{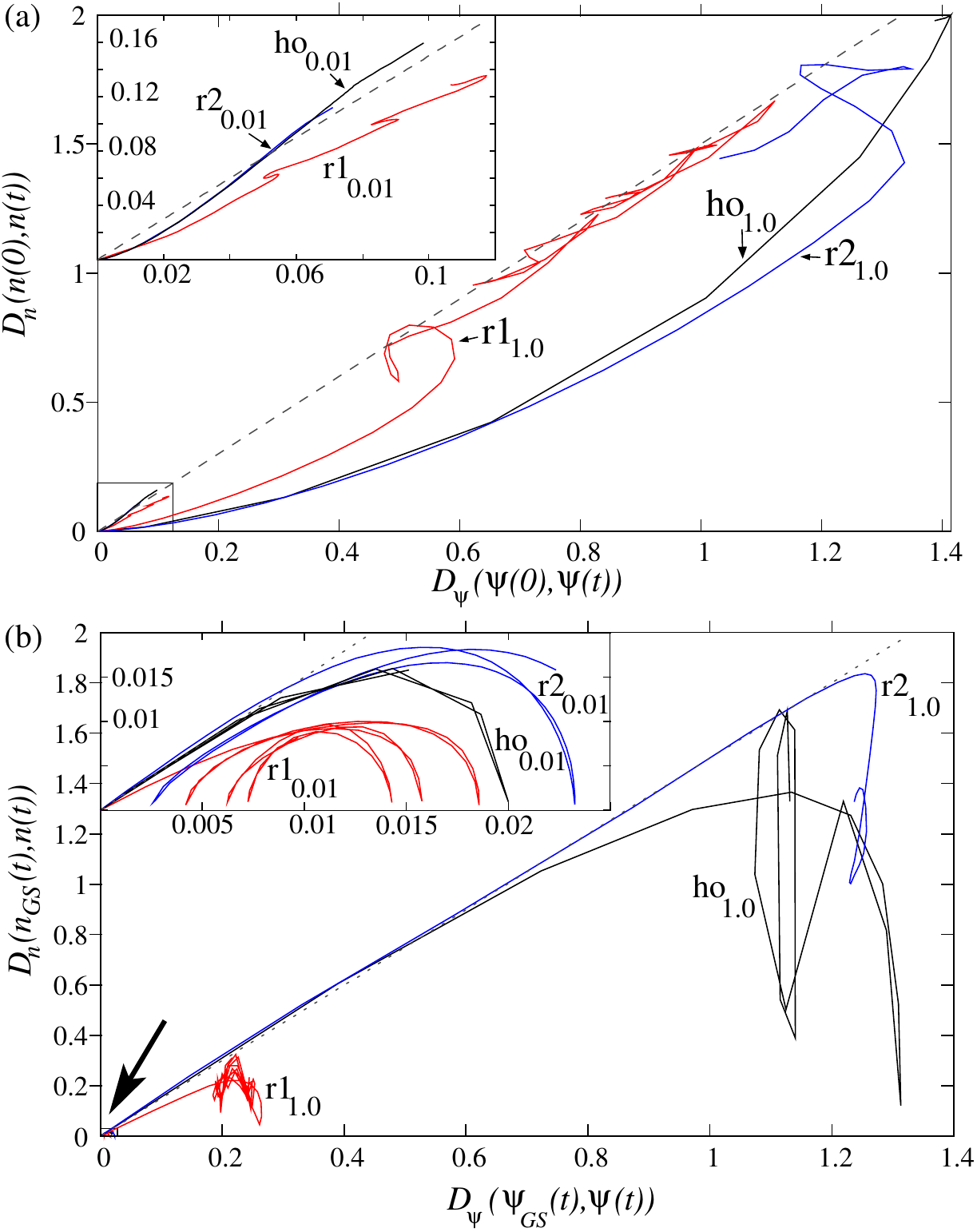}
\caption{(a) Metric distances between the initial GS and the subsequent TD state ($n$ vs.\ $\psi$): adiabatic behavior corresponds to proximity to the adiabatic (GS) line (gray dashed). Subscripts denote the value of $\epsilon(0)$.  Inset: zoom to boxed area.
(b) Distances between instantaneous GSs and TD states ($n$ vs.\ $\psi$): these should remain at the origin for exactly adiabatic evolution.  Inset: zoom to area denoted by arrow.}
\label{fig:denpsi_plot_t_plot}
\end{figure}

The first type of graph is $D_n(n(0),n(t))$ against $D_\psi(\psi(0),\psi(t))$, shown in Fig.~\ref{fig:denpsi_plot_t_plot}(a).  Here adiabaticity is identified without the direct involvement of the instantaneous GS.  It is for this graph that the gradient from Fig.~\ref{fig:gs_SHO_rand} is used.  The ratio $D_n/D_\psi$ of the distances between any two GSs is approximately given by this gradient of 1.5, and hence it can be used to characterize adiabaticity in Fig.~\ref{fig:denpsi_plot_t_plot}(a).

The systems in the inset to Fig.~\ref{fig:denpsi_plot_t_plot}(a) follow the ``adiabatic line'', showing them to be adiabatic in agreement with the corresponding $\epsilon(0)$.  Interestingly, after a transient, r1$_{1.0}$ (main panel) is also seen to follow the adiabatic line, despite the related value $\epsilon(0)=1.0$ suggesting non-adiabaticity.  In fact the metric graph shows the evolution to be be initially \textit{non}-adiabatic before returning to the adiabatic line, in agreement with $\epsilon(t)$ in Fig.~\ref{fig:gs_SHO_rand}, however $\epsilon(t)$ cannot be used to accurately determine the level of adiabaticity after a period of non-adiabatic evolution due to the use of the perfectly adiabatically evolving state.  The metrics do not suffer from this weakness and can be used to characterize a wider range of evolutions. For r2,  Fig.~\ref{fig:denpsi_plot_t_plot}(a) suggests a degree of non-adiabaticity similar to ho \footnote{These results suggest that, by combining the requirements of a dynamic ratio $D_n(n(0),n(t))/D_\psi(\psi(0),\psi(t))$ following a line, with the ``non-ergodicity'' described below, adiabatic behavior could be assessed even when the GS gradient $D_n/D_\psi$ is unknown.}.

The second type of graph is $D_n(n_{GS}(t),n(t))$ against $D_\psi(\psi_{GS}(t),\psi(t))$ [Fig.~\ref{fig:denpsi_plot_t_plot}(b)].  Here, the measure of adiabaticity comes from proximity to the origin.  We can clearly see that for $\epsilon(0) = 1.0$ (denoted in the label subscripts), ho and r2 are non-adiabatic, as $\epsilon(0)$ would suggest.  However, r1 is much closer to adiabaticity as it lies a lot closer to the origin. Systems ho and r2 display once more a similar degree of non-adiabaticity, at difference with $\epsilon(t)$ (Fig.~\ref{fig:gs_SHO_rand}, upper inset).  From this we are able to see how $\epsilon(t)$ does not always fully describe the degree of adiabaticity of the system.

We note that $D_\psi(\psi_{GS}(t),\psi(t))$ provides a \textit{quantitative} measure of the degree of adiabaticity, with $D_\psi(\psi_{GS}(t),\psi(t))=0$ indicating perfect adiabaticity and $D_\psi(\psi_{GS}(t),\psi(t))=\sqrt{N}$ corresponding to maximum non-adiabaticity [where $\psi(t)$ is either orthogonal to or completely non-overlapping with $\psi_{GS}(t)$].  This means an absolute percentage deviation of the dynamic distance from the maximum distance can be attributed at any instant in time.

This measure provides useful information beyond the degree of adiabaticity; Fig.~\ref{fig:denpsi_plot_t_plot}(b) displays oscillating ``arches'' for the adiabatic systems (inset), where ho has the clearest arches.  For ho this is seen for all values of $\epsilon(0)$ up to 1.0, where the arch is disrupted by the distortion of the harmonic well when reaching the edge of the system ($L=15$).  The frequency  of the oscillating arches is $\omega$ in the wavefunction, and $2\omega$ for the density. The random potentials also display this oscillatory behavior when adiabatic, but with a frequency not as clearly dependent on the trapping microwells' frequency.  These arches reveal a peculiar feature of the dynamics of adiabatic states: they oscillate about the instantaneous GS but never really adjust to it, maintaining this ``inertia'' no matter how slowly-varying the perturbation is.

\begin{figure}
\includegraphics[width=0.48\textwidth]{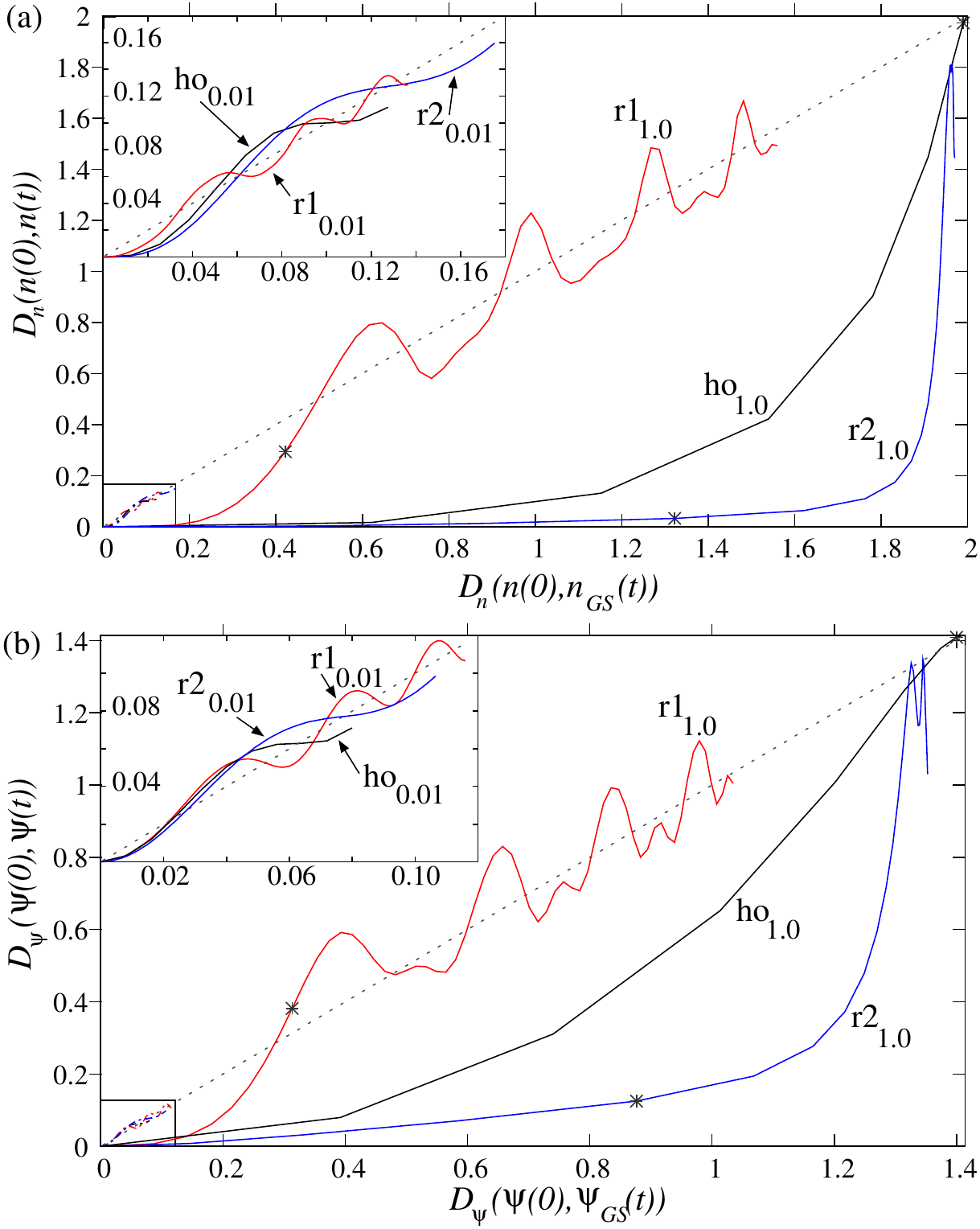}
\caption{Comparing the instantaneous GS and the TD state with the initial GS for (a) $n$ and (b) $\psi$. The black stars indicate the reference time $t_{\mathrm{ref}}$ as seen in Fig.~\ref{fig:gs_SHO_rand} (upper inset). Inset: zoom to boxed area, with adiabatic systems following the adiabatic line (dashed).}
\label{fig:den_plot}
\end{figure}

An animation for the density of ho$_{0.1}$ was produced to demonstrate the oscillations about the instantaneous ground state \cite{Skelt2017-2}. Here $\epsilon(0)=0.1$ was used as these dynamics can be seen clearer than for $\epsilon(0)=0.01$, but the oscillating arches appear in both cases.  The animation shows that the dynamic state remains superimposed to the initial ground state for a while (about 5 a.u.) after the perturbation has been applied, demonstrating inertia, before it begins to move.   This inertia of the dynamic state  gives rise to the ``ramp-up'' phase, which precedes the oscillations seen for all three families of systems (see inset of Fig.~\ref{fig:den_plot}) .

Once the dynamic state is moving, it catches up with the instantaneous ground state but due to the momentum, it continues past the instantaneous ground state until it is stopped by the potential at about 30 a.u. (where the maximum of the density has clearly overcome the minimum of the instantaneous potential) and then again at about 60a.u. (see animation \cite{Skelt2017-2}).  This causes the oscillations about the instantaneous ground state, which are seen in the insets of Figs.~\ref{fig:denpsi_plot_t_plot}(b) and \ref{fig:den_plot}.

Fig.~\ref{fig:denpsi_plot_t_plot} suggests a ``non-ergodic'' behavior for the dynamics of quantum systems in metric space, with the region above the adiabatic line remaining largely unexplored. This would imply that, on average, non-adiabaticity affects the wavefunctions more than the related densities, both when measured as a distance from the instantaneous eigenstate [Fig.~\ref{fig:denpsi_plot_t_plot} (b)]  or from the initial state [Fig.~\ref{fig:denpsi_plot_t_plot} (a)] \footnote{Preliminary results on a strongly driven, ionizing system also confirm this non-ergodicity (A. Schild, H. Gross and I. D'Amico, private communication).}.  This behavior sheds new light on the dynamic wavefunction-density mapping of TDDFT: when observed in metric space this mapping is non-ergodic; also, in contrast to the GS mapping of DFT \cite{DAmico}, it maps,  on average, close densities to less close wavefunctions. This can be partly understood by noting that distant densities must be non-overlapping (since $n$ cannot be negative) and therefore imply distant wavefunctions, whereas the converse is not true.

The third type of graph is shown in Fig.~\ref{fig:den_plot}: it focuses solely on either densities (a) or wavefunctions (b). For an adiabatic system $\psi (t) = \psi_{GS}(t)$, and so $D_\psi(\psi(0),\psi(t))=D_\psi(\psi(0),\psi_{GS}(t))$.  By comparing $D_\psi(\psi(0),\psi(t))$ with $D_\psi(\psi(0),\psi_{GS}(t))$ (or similarly with the density), the adiabaticity of the system is discerned through the proximity to the adiabatic line $y=x$. The density and wavefunction graphs are very similar, and this suggests it should be possible to determine adiabaticity using the density alone, e.g. conveniently calculated using DFT.

The systems for $\epsilon(0) = 0.01$ are indeed adiabatic and oscillate about the adiabatic line.  These oscillations  always begin below the adiabatic line: the dynamic state lags behind the instantaneous GS, in agreement with the arches seen in Fig.~\ref{fig:denpsi_plot_t_plot} (b) and  showing again the ``inertia'' felt by the dynamic system.

The region \textit{above} the adiabatic line is barely explored, once more suggesting an absence of ``ergodicity''  for the dynamics of quantum systems in metric space. For Fig.~\ref{fig:den_plot}, this may be understood using the triangle inequality obeyed by metrics, which here takes the form
$D_\psi(\psi(0),\psi(t)) \leq D_\psi(\psi_{GS}(t),\psi(t)) + D_\psi(\psi(0),\psi_{GS}(t))$ \footnote{The triangle inequality has also been used to develop limits on adiabatic time in many-body systems \cite{Lychkovskiy2017}.}. Since $D_\psi(\psi_{GS}(t),\psi(t))$ becomes smaller for increasing adiabaticity, this means that $D_\psi(\psi(0),\psi(t)) \leq D_\psi(\psi(0),\psi_{GS}(t))$ to a better and better approximation, limiting the vertical excursion of curves in Fig.~\ref{fig:den_plot}. The more adiabatic a system, the smaller the amplitude of the oscillations about the adiabatic line.  This also holds true for the density.   For ho and r2, when $\epsilon(0) = 1.0$, the region \textit{below} the adiabatic line is explored considerably, demonstrating their non-adiabatic nature.

The black stars on the $\epsilon = 1.0$ curves in Fig.~\ref{fig:den_plot} indicate $t_{\mathrm{ref}}$ (an arbitrary reference time chosen to indicate interesting dynamics) from Fig.~\ref{fig:gs_SHO_rand} (upper inset).  It is clear that r2 remains non-adiabatic at this time, however r1 has come closer to adiabaticity, and oscillates about the adiabatic line as a result of the spreading and contracting of the density in a ``breathing'' motion \footnote{Whereas Dobson's harmonic potential theorem \cite{Dobson_PhysRevLett.73.2244} shows the propensity for ``breathing'' of a time-evolving wavefunction to be suppressed in the harmonic oscillator.}.

This move towards an adiabatic regime is clearly seen in the metrics and in $\epsilon(t)$, yet the metrics, due to their non-perturbative nature, reveal a lot more about the dynamics of the system, such as the oscillations and the initial ramp-up phase due to the inertia.  They also reveal that r1$_{1.0}$ is definitely not as adiabatic as $\epsilon(t) \rightarrow 0.03$ (from Fig.~\ref{fig:gs_SHO_rand} upper inset) would suggest.

An animation of the density of r1$_{1.0}$ was produced to demonstrate this breathing motion \cite{Skelt2017-2}. From the beginning the electronic ground state is mainly confined by the asymmetric right-hand microwell and the perturbation ($-pxt$) pushes the electron closer to the confining potential as the microwell deepens.  Starting at about 30 a.u. we observe a ``breathing'' motion, with the density widening with the amplitude reducing, followed by it tightening with the amplitude increasing. This is combined with a sideway oscillation. This complex motion is caused by a combination of oscillations about the instantaneous ground state caused by inertia (similar to what mentioned previously) combined with the reflection of the wave packet by each side of the microwell in turn, an overall motion that is reminiscent of water oscillating sideway in a basin.
Each density maxima corresponds to one of the maxima of the metric oscillations observed for r1$_{1.0}$ in Fig.~\ref{fig:den_plot}(a): the higher metric maxima correspond to the density maxima close to the system boundary, while the secondary metric maxima correspond to the density maxima close to the less steep left border of the microwell.

\section{Conclusion}
In summary, we have analyzed a set of systems defined by randomly generated external potentials, using the metric-space approach to quantum mechanics. For ground states, the relationship between  $D_n$ and $D_{\psi}$ is quasi-linear over most of the possible range of values. This quasi-linearity was analytically confirmed for harmonic oscillators.
We proposed three types of metric graphs as tools to assess adiabaticity, which all agree on the character of the dynamic evolutions considered. These tools provide both \textit{quantitative and qualitative} estimates of the degree of adiabaticity in the dynamics of a quantum system, and show how the ground state linear relationship between $D_n$ and $D_{\psi}$ is related to adiabatically-evolving time-dependent systems.  All our numerical results, including additional intermediate perturbations not shown here, for these three types of graph support the conjecture that the behavior is indeed general.
We have demonstrated that the metric-space approach can be used to assess the character of the dynamics of quantum systems, in an accurate and appealingly visual way.
The metric approach studied here is also applicable to many-particle systems, for which the characterization of the degree of adiabaticity using metrics based on densities alone is particularly convenient. Our method could therefore be used to predict parameters for experiments and/or experimentally measured local densities could be used in the density metrics.  The ability to use metrics based purely on densities or wavefunctions also allows for their use in situations where only the wavefunctions or only the densities are known. An example in which the exploration of the wave function metric dynamics could be informative is the case of quantum phase transitions. Our results show that quantum dynamics, even for systems strongly far from equilibrium, appears non-ergodic in metric space. This sheds light on the density-wavefunction mapping at the core of TDDFT.
Importantly, the metric graphs do not suffer from the same limitations as the currently widely used adiabatic criterion, $\epsilon (t)$, and hence  provide a more robust indication of the degree of adiabaticity, as well as a greater insight into the system dynamics. This establishes the metric space approach to quantum mechanics as a versatile and sensitive probe of adiabaticity.

\begin{acknowledgments}
We acknowledge helpful discussions with P. Sharp, J. Wetherell, M. Hodgson and M. Herrera, and advice on his SPiDEA code from J. Wetherell. AHS acknowledges support from EPSRC; IDA acknowledges support from the Conselho Nacional de Desenvolvimento Cientfico e
Tecnologico (CNPq, Grant: PVE Processo: 401414/2014-0) and from the Royal Society (Grant no. NA140436).
\end{acknowledgments}

\bibliography{adiab_paper}

\begin{thebibliography}{43}%
\makeatletter
\providecommand \@ifxundefined [1]{%
 \@ifx{#1\undefined}
}%
\providecommand \@ifnum [1]{%
 \ifnum #1\expandafter \@firstoftwo
 \else \expandafter \@secondoftwo
 \fi
}%
\providecommand \@ifx [1]{%
 \ifx #1\expandafter \@firstoftwo
 \else \expandafter \@secondoftwo
 \fi
}%
\providecommand \natexlab [1]{#1}%
\providecommand \enquote  [1]{``#1''}%
\providecommand \bibnamefont  [1]{#1}%
\providecommand \bibfnamefont [1]{#1}%
\providecommand \citenamefont [1]{#1}%
\providecommand \href@noop [0]{\@secondoftwo}%
\providecommand \href [0]{\begingroup \@sanitize@url \@href}%
\providecommand \@href[1]{\@@startlink{#1}\@@href}%
\providecommand \@@href[1]{\endgroup#1\@@endlink}%
\providecommand \@sanitize@url [0]{\catcode `\\12\catcode `\$12\catcode
  `\&12\catcode `\#12\catcode `\^12\catcode `\_12\catcode `\%12\relax}%
\providecommand \@@startlink[1]{}%
\providecommand \@@endlink[0]{}%
\providecommand \url  [0]{\begingroup\@sanitize@url \@url }%
\providecommand \@url [1]{\endgroup\@href {#1}{\urlprefix }}%
\providecommand \urlprefix  [0]{URL }%
\providecommand \Eprint [0]{\href }%
\providecommand \doibase [0]{http://dx.doi.org/}%
\providecommand \selectlanguage [0]{\@gobble}%
\providecommand \bibinfo  [0]{\@secondoftwo}%
\providecommand \bibfield  [0]{\@secondoftwo}%
\providecommand \translation [1]{[#1]}%
\providecommand \BibitemOpen [0]{}%
\providecommand \bibitemStop [0]{}%
\providecommand \bibitemNoStop [0]{.\EOS\space}%
\providecommand \EOS [0]{\spacefactor3000\relax}%
\providecommand \BibitemShut  [1]{\csname bibitem#1\endcsname}%
\let\auto@bib@innerbib\@empty
\bibitem [{\citenamefont {Kosloff}(2013)}]{Kosloff2013}%
  \BibitemOpen
  \bibfield  {author} {\bibinfo {author} {\bibfnamefont {R.}~\bibnamefont
  {Kosloff}},\ }\bibfield  {title} {\enquote {\bibinfo {title} {Quantum
  thermodynamics: A dynamical viewpoint},}\ }\href@noop {} {\bibfield
  {journal} {\bibinfo  {journal} {Entropy}\ }\textbf {\bibinfo {volume} {15}},\
  \bibinfo {pages} {2100} (\bibinfo {year} {2013})}\BibitemShut {NoStop}%
\bibitem [{\citenamefont {Vinjanampathy}\ and\ \citenamefont
  {Anders}(2016)}]{Vinjanampathy2016}%
  \BibitemOpen
  \bibfield  {author} {\bibinfo {author} {\bibfnamefont {S.}~\bibnamefont
  {Vinjanampathy}}\ and\ \bibinfo {author} {\bibfnamefont {J.}~\bibnamefont
  {Anders}},\ }\bibfield  {title} {\enquote {\bibinfo {title} {Quantum
  thermodynamics},}\ }\href@noop {} {\bibfield  {journal} {\bibinfo  {journal}
  {Contemp. Phys.}\ }\textbf {\bibinfo {volume} {57}},\ \bibinfo {pages} {545}
  (\bibinfo {year} {2016})}\BibitemShut {NoStop}%
\bibitem [{\citenamefont {Millen}\ and\ \citenamefont
  {Xuereb}(2016)}]{Millen2016}%
  \BibitemOpen
  \bibfield  {author} {\bibinfo {author} {\bibfnamefont {J.}~\bibnamefont
  {Millen}}\ and\ \bibinfo {author} {\bibfnamefont {A.}~\bibnamefont
  {Xuereb}},\ }\bibfield  {title} {\enquote {\bibinfo {title} {Perspective on
  quantum thermodynamics},}\ }\href@noop {} {\bibfield  {journal} {\bibinfo
  {journal} {New J. Phys.}\ }\textbf {\bibinfo {volume} {18}},\ \bibinfo
  {pages} {011002} (\bibinfo {year} {2016})}\BibitemShut {NoStop}%
\bibitem [{\citenamefont {Shevchenko}\ \emph {et~al.}(2010)\citenamefont
  {Shevchenko}, \citenamefont {Ashhab},\ and\ \citenamefont
  {Nori}}]{Shevchenko2010}%
  \BibitemOpen
  \bibfield  {author} {\bibinfo {author} {\bibfnamefont {S.~N.}\ \bibnamefont
  {Shevchenko}}, \bibinfo {author} {\bibfnamefont {S.}~\bibnamefont {Ashhab}},
  \ and\ \bibinfo {author} {\bibfnamefont {Franco}\ \bibnamefont {Nori}},\
  }\bibfield  {title} {\enquote {\bibinfo {title}
  {Landau–zener–stückelberg interferometry},}\ }\href@noop {} {\bibfield
  {journal} {\bibinfo  {journal} {Physics Reports}\ }\textbf {\bibinfo {volume}
  {492}},\ \bibinfo {pages} {1} (\bibinfo {year} {2010})}\BibitemShut {NoStop}%
\bibitem [{\citenamefont {Berry}(1984)}]{Berry1984}%
  \BibitemOpen
  \bibfield  {author} {\bibinfo {author} {\bibfnamefont {M.~V.}\ \bibnamefont
  {Berry}},\ }\bibfield  {title} {\enquote {\bibinfo {title} {Quantal phase
  factors accompanying adiabatic changes},}\ }\href@noop {} {\bibfield
  {journal} {\bibinfo  {journal} {Proc. R. Soc. A}\ }\textbf {\bibinfo {volume}
  {392}},\ \bibinfo {pages} {45} (\bibinfo {year} {1984})}\BibitemShut
  {NoStop}%
\bibitem [{\citenamefont {Laughlin}(1981)}]{Laughlin1981}%
  \BibitemOpen
  \bibfield  {author} {\bibinfo {author} {\bibfnamefont {R.~B.}\ \bibnamefont
  {Laughlin}},\ }\bibfield  {title} {\enquote {\bibinfo {title} {Quantized hall
  conductivity in two dimensions},}\ }\href@noop {} {\bibfield  {journal}
  {\bibinfo  {journal} {Phys. Rev. B}\ }\textbf {\bibinfo {volume} {23}},\
  \bibinfo {pages} {5632} (\bibinfo {year} {1981})}\BibitemShut {NoStop}%
\bibitem [{\citenamefont {Bergmann}\ \emph {et~al.}(1998)\citenamefont
  {Bergmann}, \citenamefont {Theuer},\ and\ \citenamefont
  {Shore}}]{Bergmann1998}%
  \BibitemOpen
  \bibfield  {author} {\bibinfo {author} {\bibfnamefont {K.}~\bibnamefont
  {Bergmann}}, \bibinfo {author} {\bibfnamefont {H.}~\bibnamefont {Theuer}}, \
  and\ \bibinfo {author} {\bibfnamefont {B.}~\bibnamefont {Shore}},\ }\bibfield
   {title} {\enquote {\bibinfo {title} {Coherent population transfer among
  quantum states of atoms and molecules},}\ }\href@noop {} {\bibfield
  {journal} {\bibinfo  {journal} {Rev. Mod. Phys.}\ }\textbf {\bibinfo {volume}
  {70}},\ \bibinfo {pages} {1003} (\bibinfo {year} {1998})}\BibitemShut
  {NoStop}%
\bibitem [{\citenamefont {Vitanov}\ \emph {et~al.}(2001)\citenamefont
  {Vitanov}, \citenamefont {Halfmann}, \citenamefont {Shore},\ and\
  \citenamefont {Bergmann}}]{Vitanov2001}%
  \BibitemOpen
  \bibfield  {author} {\bibinfo {author} {\bibfnamefont {N.~V.}\ \bibnamefont
  {Vitanov}}, \bibinfo {author} {\bibfnamefont {T.}~\bibnamefont {Halfmann}},
  \bibinfo {author} {\bibfnamefont {B.~W.}\ \bibnamefont {Shore}}, \ and\
  \bibinfo {author} {\bibfnamefont {K.}~\bibnamefont {Bergmann}},\ }\bibfield
  {title} {\enquote {\bibinfo {title} {Laser-induced population transfer by
  adiabatic passage techniques},}\ }\href@noop {} {\bibfield  {journal}
  {\bibinfo  {journal} {Annu. Rev. Phys. Chem.}\ }\textbf {\bibinfo {volume}
  {52}},\ \bibinfo {pages} {763} (\bibinfo {year} {2001})}\BibitemShut
  {NoStop}%
\bibitem [{\citenamefont {Greentree}\ \emph {et~al.}(2004)\citenamefont
  {Greentree}, \citenamefont {Cole}, \citenamefont {Hamilton},\ and\
  \citenamefont {Hollenberg}}]{Greentree2004}%
  \BibitemOpen
  \bibfield  {author} {\bibinfo {author} {\bibfnamefont {A.~D.}\ \bibnamefont
  {Greentree}}, \bibinfo {author} {\bibfnamefont {J.~H.}\ \bibnamefont {Cole}},
  \bibinfo {author} {\bibfnamefont {A.~R.}\ \bibnamefont {Hamilton}}, \ and\
  \bibinfo {author} {\bibfnamefont {L.~C.~L.}\ \bibnamefont {Hollenberg}},\
  }\bibfield  {title} {\enquote {\bibinfo {title} {Coherent electronic transfer
  in quantum dot systems using adiabatic passage},}\ }\href@noop {} {\bibfield
  {journal} {\bibinfo  {journal} {Phys. Rev. B}\ }\textbf {\bibinfo {volume}
  {70}},\ \bibinfo {pages} {235317} (\bibinfo {year} {2004})}\BibitemShut
  {NoStop}%
\bibitem [{\citenamefont {Farhi}\ \emph {et~al.}(2001)\citenamefont {Farhi},
  \citenamefont {Goldstone}, \citenamefont {Gutmann}, \citenamefont {Lapan},
  \citenamefont {Lundgren},\ and\ \citenamefont {Preda}}]{Fahri2001}%
  \BibitemOpen
  \bibfield  {author} {\bibinfo {author} {\bibfnamefont {E.}~\bibnamefont
  {Farhi}}, \bibinfo {author} {\bibfnamefont {J.}~\bibnamefont {Goldstone}},
  \bibinfo {author} {\bibfnamefont {S.}~\bibnamefont {Gutmann}}, \bibinfo
  {author} {\bibfnamefont {J.}~\bibnamefont {Lapan}}, \bibinfo {author}
  {\bibfnamefont {A.}~\bibnamefont {Lundgren}}, \ and\ \bibinfo {author}
  {\bibfnamefont {D.}~\bibnamefont {Preda}},\ }\bibfield  {title} {\enquote
  {\bibinfo {title} {A quantum adiabatic evolution algorithm applied to random
  instances of an np-complete problem},}\ }\href@noop {} {\bibfield  {journal}
  {\bibinfo  {journal} {Science}\ }\textbf {\bibinfo {volume} {292}},\ \bibinfo
  {pages} {472} (\bibinfo {year} {2001})}\BibitemShut {NoStop}%
\bibitem [{\citenamefont {Roland}\ and\ \citenamefont
  {Cerf}(2002)}]{Roland2002}%
  \BibitemOpen
  \bibfield  {author} {\bibinfo {author} {\bibfnamefont {J.}~\bibnamefont
  {Roland}}\ and\ \bibinfo {author} {\bibfnamefont {N.~J.}\ \bibnamefont
  {Cerf}},\ }\bibfield  {title} {\enquote {\bibinfo {title} {Quantum search by
  local adiabatic evolution},}\ }\href@noop {} {\bibfield  {journal} {\bibinfo
  {journal} {Phys. Rev. A}\ }\textbf {\bibinfo {volume} {65}},\ \bibinfo
  {pages} {042308} (\bibinfo {year} {2002})}\BibitemShut {NoStop}%
\bibitem [{\citenamefont {Das}\ \emph {et~al.}(2002)\citenamefont {Das},
  \citenamefont {Kobes},\ and\ \citenamefont {Kunstatter}}]{Das2002}%
  \BibitemOpen
  \bibfield  {author} {\bibinfo {author} {\bibfnamefont {S.}~\bibnamefont
  {Das}}, \bibinfo {author} {\bibfnamefont {R.}~\bibnamefont {Kobes}}, \ and\
  \bibinfo {author} {\bibfnamefont {G.}~\bibnamefont {Kunstatter}},\ }\bibfield
   {title} {\enquote {\bibinfo {title} {Adiabatic quantum computation and
  {D}eutsch's algorithm},}\ }\href@noop {} {\bibfield  {journal} {\bibinfo
  {journal} {Phys. Rev. A}\ }\textbf {\bibinfo {volume} {65}},\ \bibinfo
  {pages} {062310} (\bibinfo {year} {2002})}\BibitemShut {NoStop}%
\bibitem [{\citenamefont {Albash}\ and\ \citenamefont {Lidar}()}]{Albash2016}%
  \BibitemOpen
  \bibfield  {author} {\bibinfo {author} {\bibfnamefont {T.}~\bibnamefont
  {Albash}}\ and\ \bibinfo {author} {\bibfnamefont {D.~A.}\ \bibnamefont
  {Lidar}},\ }\href@noop {} {\enquote {\bibinfo {title} {Adiabatic quantum
  computing},}\ }\Eprint {http://arxiv.org/abs/arXiv:1611.04471}
  {arXiv:1611.04471} \BibitemShut {NoStop}%
\bibitem [{\citenamefont {Plastina}\ \emph {et~al.}(2014)\citenamefont
  {Plastina}, \citenamefont {Alecce}, \citenamefont {Apollaro}, \citenamefont
  {Falcone}, \citenamefont {Francica}, \citenamefont {Galve}, \citenamefont
  {Gullo},\ and\ \citenamefont {Zambrini}}]{Plastina2014}%
  \BibitemOpen
  \bibfield  {author} {\bibinfo {author} {\bibfnamefont {F.}~\bibnamefont
  {Plastina}}, \bibinfo {author} {\bibfnamefont {A.}~\bibnamefont {Alecce}},
  \bibinfo {author} {\bibfnamefont {T.~J.~G.}\ \bibnamefont {Apollaro}},
  \bibinfo {author} {\bibfnamefont {G.}~\bibnamefont {Falcone}}, \bibinfo
  {author} {\bibfnamefont {G.}~\bibnamefont {Francica}}, \bibinfo {author}
  {\bibfnamefont {F.}~\bibnamefont {Galve}}, \bibinfo {author} {\bibfnamefont
  {N.~Lo}\ \bibnamefont {Gullo}}, \ and\ \bibinfo {author} {\bibfnamefont
  {R.}~\bibnamefont {Zambrini}},\ }\bibfield  {title} {\enquote {\bibinfo
  {title} {Irreversible work and inner friction in quantum thermodynamic
  processes},}\ }\href@noop {} {\bibfield  {journal} {\bibinfo  {journal}
  {Phys. Rev. Lett.}\ }\textbf {\bibinfo {volume} {113}},\ \bibinfo {pages}
  {260601} (\bibinfo {year} {2014})}\BibitemShut {NoStop}%
\bibitem [{\citenamefont {Johnson}\ \emph {et~al.}(2011)\citenamefont
  {Johnson}, \citenamefont {Amin}, \citenamefont {Gildert}, \citenamefont
  {Lanting}, \citenamefont {Hamze}, \citenamefont {Dickson}, \citenamefont
  {Harris}, \citenamefont {Berkley}, \citenamefont {Johansson}, \citenamefont
  {Bunyk} \emph {et~al.}}]{Johnson2011}%
  \BibitemOpen
  \bibfield  {author} {\bibinfo {author} {\bibfnamefont {M.~W.}\ \bibnamefont
  {Johnson}}, \bibinfo {author} {\bibfnamefont {M.~H.~S.}\ \bibnamefont
  {Amin}}, \bibinfo {author} {\bibfnamefont {S.}~\bibnamefont {Gildert}},
  \bibinfo {author} {\bibfnamefont {T.}~\bibnamefont {Lanting}}, \bibinfo
  {author} {\bibfnamefont {F.}~\bibnamefont {Hamze}}, \bibinfo {author}
  {\bibfnamefont {N.}~\bibnamefont {Dickson}}, \bibinfo {author} {\bibfnamefont
  {R.}~\bibnamefont {Harris}}, \bibinfo {author} {\bibfnamefont {A.~J.}\
  \bibnamefont {Berkley}}, \bibinfo {author} {\bibfnamefont {J.}~\bibnamefont
  {Johansson}}, \bibinfo {author} {\bibfnamefont {P.}~\bibnamefont {Bunyk}},
  \emph {et~al.},\ }\bibfield  {title} {\enquote {\bibinfo {title} {Quantum
  annealing with manufactured spin},}\ }\href@noop {} {\bibfield  {journal}
  {\bibinfo  {journal} {Nature}\ }\textbf {\bibinfo {volume} {473}},\ \bibinfo
  {pages} {194} (\bibinfo {year} {2011})}\BibitemShut {NoStop}%
\bibitem [{\citenamefont {Gibney}(2017)}]{Gibney2017}%
  \BibitemOpen
  \bibfield  {author} {\bibinfo {author} {\bibfnamefont {E.}~\bibnamefont
  {Gibney}},\ }\bibfield  {title} {\enquote {\bibinfo {title} {D-wave upgrade:
  How scientists are using the world's most controversial quantum computer},}\
  }\href@noop {} {\bibfield  {journal} {\bibinfo  {journal} {Nature}\ }\textbf
  {\bibinfo {volume} {541}},\ \bibinfo {pages} {447} (\bibinfo {year}
  {2017})}\BibitemShut {NoStop}%
\bibitem [{\citenamefont {Born}\ and\ \citenamefont {Fock}(1928)}]{BornFock}%
  \BibitemOpen
  \bibfield  {author} {\bibinfo {author} {\bibfnamefont {M.}~\bibnamefont
  {Born}}\ and\ \bibinfo {author} {\bibfnamefont {V.~A.}\ \bibnamefont
  {Fock}},\ }\bibfield  {title} {\enquote {\bibinfo {title} {Beweis des
  adiabatensatzes},}\ }\href@noop {} {\bibfield  {journal} {\bibinfo  {journal}
  {Z. Phys. A}\ }\textbf {\bibinfo {volume} {51}},\ \bibinfo {pages}
  {165–180} (\bibinfo {year} {1928})}\BibitemShut {NoStop}%
\bibitem [{\citenamefont {Tong}\ \emph {et~al.}(2005)\citenamefont {Tong},
  \citenamefont {Singh}, \citenamefont {Kwek},\ and\ \citenamefont
  {Oh}}]{Tong2005}%
  \BibitemOpen
  \bibfield  {author} {\bibinfo {author} {\bibfnamefont {D.~M.}\ \bibnamefont
  {Tong}}, \bibinfo {author} {\bibfnamefont {K.}~\bibnamefont {Singh}},
  \bibinfo {author} {\bibfnamefont {L.~C.}\ \bibnamefont {Kwek}}, \ and\
  \bibinfo {author} {\bibfnamefont {C.~H.}\ \bibnamefont {Oh}},\ }\bibfield
  {title} {\enquote {\bibinfo {title} {Quantitative conditions do not guarantee
  the validity of the adiabatic approximation},}\ }\href@noop {} {\bibfield
  {journal} {\bibinfo  {journal} {Phys. Rev. Lett.}\ }\textbf {\bibinfo
  {volume} {95}},\ \bibinfo {pages} {110407} (\bibinfo {year}
  {2005})}\BibitemShut {NoStop}%
\bibitem [{\citenamefont {Marzlin}\ and\ \citenamefont
  {Sanders}(2004)}]{Marzlin2004}%
  \BibitemOpen
  \bibfield  {author} {\bibinfo {author} {\bibfnamefont {K.-P.}\ \bibnamefont
  {Marzlin}}\ and\ \bibinfo {author} {\bibfnamefont {B.~C.}\ \bibnamefont
  {Sanders}},\ }\bibfield  {title} {\enquote {\bibinfo {title} {Inconsistency
  in the application of the adiabatic theorem},}\ }\href@noop {} {\bibfield
  {journal} {\bibinfo  {journal} {Phys. Rev. Lett.}\ }\textbf {\bibinfo
  {volume} {93}},\ \bibinfo {pages} {160408} (\bibinfo {year}
  {2004})}\BibitemShut {NoStop}%
\bibitem [{\citenamefont {Ortigoso}(2012)}]{Ortigoso2012}%
  \BibitemOpen
  \bibfield  {author} {\bibinfo {author} {\bibfnamefont {J.}~\bibnamefont
  {Ortigoso}},\ }\bibfield  {title} {\enquote {\bibinfo {title} {Quantum
  adiabatic theorem in light of the {M}arzlin-{S}anders inconsistency},}\
  }\href@noop {} {\bibfield  {journal} {\bibinfo  {journal} {Phys. Rev. A}\
  }\textbf {\bibinfo {volume} {86}},\ \bibinfo {pages} {032121} (\bibinfo
  {year} {2012})}\BibitemShut {NoStop}%
\bibitem [{\citenamefont {Comparat}(2009)}]{Comparat2009}%
  \BibitemOpen
  \bibfield  {author} {\bibinfo {author} {\bibfnamefont {D.}~\bibnamefont
  {Comparat}},\ }\bibfield  {title} {\enquote {\bibinfo {title} {General
  conditions for quantum adiabatic evolution},}\ }\href@noop {} {\bibfield
  {journal} {\bibinfo  {journal} {Phys. Rev. A}\ }\textbf {\bibinfo {volume}
  {80}},\ \bibinfo {pages} {012106} (\bibinfo {year} {2009})}\BibitemShut
  {NoStop}%
\bibitem [{\citenamefont {Li}(2016)}]{Li2016}%
  \BibitemOpen
  \bibfield  {author} {\bibinfo {author} {\bibfnamefont {D.}~\bibnamefont
  {Li}},\ }\bibfield  {title} {\enquote {\bibinfo {title} {Invalidity of the
  quantitative adiabatic condition and general conditions for adiabatic
  approximations},}\ }\href@noop {} {\bibfield  {journal} {\bibinfo  {journal}
  {Laser Phys. Lett.}\ }\textbf {\bibinfo {volume} {13}},\ \bibinfo {pages}
  {055203} (\bibinfo {year} {2016})}\BibitemShut {NoStop}%
\bibitem [{\citenamefont {Du}\ \emph {et~al.}(2008)\citenamefont {Du},
  \citenamefont {Hu}, \citenamefont {Wang}, \citenamefont {Wu}, \citenamefont
  {Zhao},\ and\ \citenamefont {Suter}}]{Du2008}%
  \BibitemOpen
  \bibfield  {author} {\bibinfo {author} {\bibfnamefont {Jiangfeng}\
  \bibnamefont {Du}}, \bibinfo {author} {\bibfnamefont {Lingzhi}\ \bibnamefont
  {Hu}}, \bibinfo {author} {\bibfnamefont {Ya}~\bibnamefont {Wang}}, \bibinfo
  {author} {\bibfnamefont {Jianda}\ \bibnamefont {Wu}}, \bibinfo {author}
  {\bibfnamefont {Meisheng}\ \bibnamefont {Zhao}}, \ and\ \bibinfo {author}
  {\bibfnamefont {Dieter}\ \bibnamefont {Suter}},\ }\bibfield  {title}
  {\enquote {\bibinfo {title} {Experimental study of the validity of
  quantitative conditions in the quantum adiabatic theorem},}\ }\href@noop {}
  {\bibfield  {journal} {\bibinfo  {journal} {Phys. Rev. Lett.}\ }\textbf
  {\bibinfo {volume} {101}},\ \bibinfo {pages} {060403} (\bibinfo {year}
  {2008})}\BibitemShut {NoStop}%
\bibitem [{\citenamefont {Jansen}\ \emph {et~al.}(2007)\citenamefont {Jansen},
  \citenamefont {Ruskai},\ and\ \citenamefont {Seiler}}]{Jansen2007}%
  \BibitemOpen
  \bibfield  {author} {\bibinfo {author} {\bibfnamefont {Sabine}\ \bibnamefont
  {Jansen}}, \bibinfo {author} {\bibfnamefont {Mary-Beth}\ \bibnamefont
  {Ruskai}}, \ and\ \bibinfo {author} {\bibfnamefont {Ruedi}\ \bibnamefont
  {Seiler}},\ }\bibfield  {title} {\enquote {\bibinfo {title} {Bounds for the
  adiabatic approximation with applications to quantum computation},}\
  }\href@noop {} {\bibfield  {journal} {\bibinfo  {journal} {Journal of
  Mathematical Physics}\ }\textbf {\bibinfo {volume} {48}},\ \bibinfo {pages}
  {102111} (\bibinfo {year} {2007})}\BibitemShut {NoStop}%
\bibitem [{\citenamefont {Amin}(2009)}]{Amin2009}%
  \BibitemOpen
  \bibfield  {author} {\bibinfo {author} {\bibfnamefont {M.~H.~S.}\
  \bibnamefont {Amin}},\ }\bibfield  {title} {\enquote {\bibinfo {title}
  {Consistency of the adiabatic theorem},}\ }\href@noop {} {\bibfield
  {journal} {\bibinfo  {journal} {Phys. Rev. Lett.}\ }\textbf {\bibinfo
  {volume} {102}},\ \bibinfo {pages} {220401} (\bibinfo {year}
  {2009})}\BibitemShut {NoStop}%
\bibitem [{\citenamefont {D'Amico}\ \emph {et~al.}(2011)\citenamefont
  {D'Amico}, \citenamefont {Coe}, \citenamefont {Fran\c{c}a},\ and\
  \citenamefont {Capelle}}]{DAmico}%
  \BibitemOpen
  \bibfield  {author} {\bibinfo {author} {\bibfnamefont {I.}~\bibnamefont
  {D'Amico}}, \bibinfo {author} {\bibfnamefont {J.~P.}\ \bibnamefont {Coe}},
  \bibinfo {author} {\bibfnamefont {V.~V.}\ \bibnamefont {Fran\c{c}a}}, \ and\
  \bibinfo {author} {\bibfnamefont {K}~\bibnamefont {Capelle}},\ }\bibfield
  {title} {\enquote {\bibinfo {title} {Quantum mechanics in metric space: Wave
  functions and their densities},}\ }\href@noop {} {\bibfield  {journal}
  {\bibinfo  {journal} {Phys. Rev. Lett.}\ }\textbf {\bibinfo {volume} {106}},\
  \bibinfo {pages} {050401} (\bibinfo {year} {2011})}\BibitemShut {NoStop}%
\bibitem [{\citenamefont {Sharp}\ and\ \citenamefont {D’Amico}(2014)}]{S&D}%
  \BibitemOpen
  \bibfield  {author} {\bibinfo {author} {\bibfnamefont {P.~M.}\ \bibnamefont
  {Sharp}}\ and\ \bibinfo {author} {\bibfnamefont {I.}~\bibnamefont
  {D’Amico}},\ }\bibfield  {title} {\enquote {\bibinfo {title} {Metric space
  formulation of quantum mechanical conservation laws},}\ }\href@noop {}
  {\bibfield  {journal} {\bibinfo  {journal} {Phys. Rev. B}\ }\textbf {\bibinfo
  {volume} {89}},\ \bibinfo {pages} {115137} (\bibinfo {year}
  {2014})}\BibitemShut {NoStop}%
\bibitem [{\citenamefont {Runge}\ and\ \citenamefont {Gross}(1984)}]{Runge}%
  \BibitemOpen
  \bibfield  {author} {\bibinfo {author} {\bibfnamefont {E.}~\bibnamefont
  {Runge}}\ and\ \bibinfo {author} {\bibfnamefont {E.~K.~U.}\ \bibnamefont
  {Gross}},\ }\bibfield  {title} {\enquote {\bibinfo {title}
  {Density-functional theory for time-dependent systems},}\ }\href@noop {}
  {\bibfield  {journal} {\bibinfo  {journal} {Phys. Rev. Lett.}\ }\textbf
  {\bibinfo {volume} {52}},\ \bibinfo {pages} {997} (\bibinfo {year}
  {1984})}\BibitemShut {NoStop}%
\bibitem [{\citenamefont {Sharp}\ and\ \citenamefont {D’Amico}(2015)}]{S&D2}%
  \BibitemOpen
  \bibfield  {author} {\bibinfo {author} {\bibfnamefont {P.~M.}\ \bibnamefont
  {Sharp}}\ and\ \bibinfo {author} {\bibfnamefont {I.}~\bibnamefont
  {D’Amico}},\ }\bibfield  {title} {\enquote {\bibinfo {title} {Metric space
  analysis of systems immersed in a magnetic field},}\ }\href@noop {}
  {\bibfield  {journal} {\bibinfo  {journal} {Phys. Rev. A}\ }\textbf {\bibinfo
  {volume} {92}},\ \bibinfo {pages} {032509} (\bibinfo {year}
  {2015})}\BibitemShut {NoStop}%
\bibitem [{Note1()}]{Note1}%
  \BibitemOpen
  \bibinfo {note} {We use atomic units, ${\mathchar '26\mkern
  -9muh}=m=1$}\BibitemShut {NoStop}%
\bibitem [{Note2()}]{Note2}%
  \BibitemOpen
  \bibinfo {note} {\label {SPiDEA}J. Wetherell, unpublished; subsequently
  incorporated into the iDEA code suite \cite {Hodgson}}\BibitemShut {NoStop}%
\bibitem [{Note3()}]{Note3}%
  \BibitemOpen
  \bibinfo {note} {Preliminary results for random potentials with $N$=2 also
  show a quasi-linear relationship.}\BibitemShut {Stop}%
\bibitem [{Note4()}]{Note4}%
  \BibitemOpen
  \bibinfo {note} {For the evolution of GSs this quasi-linear relationship
  indicates both equilibrium and adiabaticity; Ref.~\cite {S&D2} suggests that
  a similar relationship may hold also for excited states, hence the proposed
  method could be extended to any eigenstate}\BibitemShut {NoStop}%
\bibitem [{Note5()}]{Note5}%
  \BibitemOpen
  \bibinfo {note} {Euclidean distances between wavefunctions have previously
  been used \cite {Li2016} to study the validity of $\epsilon (t)$, but are
  inappropriately sensitive to a physically-irrelevant overall phase-change of
  the state; the metrics used here are tailored to avoid this shortcoming \cite
  {DAmico}}\BibitemShut {NoStop}%
\bibitem [{Ske()}]{Skelt2017-2}%
  \BibitemOpen
  \href@noop {} {}\bibinfo {note} {{S}ee Supplemental Material for videos of
  the variety of dynamics, the breathing dynamics and explanations of the
  inertia.}\BibitemShut {Stop}%
\bibitem [{Note6()}]{Note6}%
  \BibitemOpen
  \bibinfo {note} {For $\epsilon (0)=0.01$, the values of $p$ are 2.530, 0.15
  and 0.025 for the ho, r1 and r2 systems, respectively, while for $\epsilon
  (0)=1.0$ the values of $p$ are 100 times greater.}\BibitemShut {Stop}%
\bibitem [{Note7()}]{Note7}%
  \BibitemOpen
  \bibinfo {note} {These results suggest that, by combining the requirements of
  a dynamic ratio $D_n(n(0),n(t))/D_\psi (\psi (0),\psi (t))$ following a line,
  with the ``non-ergodicity'' described below, adiabatic behavior could be
  assessed even when the GS gradient $D_n/D_\psi $ is unknown.}\BibitemShut
  {Stop}%
\bibitem [{Note8()}]{Note8}%
  \BibitemOpen
  \bibinfo {note} {Preliminary results on a strongly driven, ionizing system
  also confirm this non-ergodicity (A. Schild, H. Gross and I. D'Amico, private
  communication).}\BibitemShut {Stop}%
\bibitem [{Note9()}]{Note9}%
  \BibitemOpen
  \bibinfo {note} {The triangle inequality has also been used to develop limits
  on adiabatic time in many-body systems \cite {Lychkovskiy2017}.}\BibitemShut
  {Stop}%
\bibitem [{Note10()}]{Note10}%
  \BibitemOpen
  \bibinfo {note} {Whereas Dobson's harmonic potential theorem \cite
  {Dobson_PhysRevLett.73.2244} shows the propensity for ``breathing'' of a
  time-evolving wavefunction to be suppressed in the harmonic
  oscillator.}\BibitemShut {Stop}%
\bibitem [{\citenamefont {Hodgson}\ \emph {et~al.}(2013)\citenamefont
  {Hodgson}, \citenamefont {Ramsden}, \citenamefont {Chapman}, \citenamefont
  {Lillystone},\ and\ \citenamefont {Godby}}]{Hodgson}%
  \BibitemOpen
  \bibfield  {author} {\bibinfo {author} {\bibfnamefont {M.~J.~P.}\
  \bibnamefont {Hodgson}}, \bibinfo {author} {\bibfnamefont {J.~D.}\
  \bibnamefont {Ramsden}}, \bibinfo {author} {\bibfnamefont {J.~B.~J.}\
  \bibnamefont {Chapman}}, \bibinfo {author} {\bibfnamefont {P.}~\bibnamefont
  {Lillystone}}, \ and\ \bibinfo {author} {\bibfnamefont {R.~W.}\ \bibnamefont
  {Godby}},\ }\bibfield  {title} {\enquote {\bibinfo {title} {Exact
  time-dependent density-functional potentials for strongly correlated
  tunneling electrons},}\ }\href@noop {} {\bibfield  {journal} {\bibinfo
  {journal} {Phys. Rev. B}\ }\textbf {\bibinfo {volume} {88}},\ \bibinfo
  {pages} {241102(R)} (\bibinfo {year} {2013})}\BibitemShut {NoStop}%
\bibitem [{\citenamefont {Lychkovskiy}\ \emph {et~al.}(2017)\citenamefont
  {Lychkovskiy}, \citenamefont {Gamayun},\ and\ \citenamefont
  {Cheianov}}]{Lychkovskiy2017}%
  \BibitemOpen
  \bibfield  {author} {\bibinfo {author} {\bibfnamefont {Oleg}\ \bibnamefont
  {Lychkovskiy}}, \bibinfo {author} {\bibfnamefont {Oleksandr}\ \bibnamefont
  {Gamayun}}, \ and\ \bibinfo {author} {\bibfnamefont {Vadim}\ \bibnamefont
  {Cheianov}},\ }\bibfield  {title} {\enquote {\bibinfo {title} {Time scale for
  adiabaticity breakdown in driven many-body systems and orthogonality
  catastrophe},}\ }\href@noop {} {\bibfield  {journal} {\bibinfo  {journal}
  {Phys. Rev. Lett.}\ }\textbf {\bibinfo {volume} {119}},\ \bibinfo {pages}
  {200401} (\bibinfo {year} {2017})}\BibitemShut {NoStop}%
\bibitem [{\citenamefont {Dobson}(1994)}]{Dobson_PhysRevLett.73.2244}%
  \BibitemOpen
  \bibfield  {author} {\bibinfo {author} {\bibfnamefont {John~F.}\ \bibnamefont
  {Dobson}},\ }\bibfield  {title} {\enquote {\bibinfo {title}
  {Harmonic-potential theorem: Implications for approximate many-body
  theories},}\ }\href {\doibase 10.1103/PhysRevLett.73.2244} {\bibfield
  {journal} {\bibinfo  {journal} {Phys. Rev. Lett.}\ }\textbf {\bibinfo
  {volume} {73}},\ \bibinfo {pages} {2244--2247} (\bibinfo {year}
  {1994})}\BibitemShut {NoStop}%
\end{thebibliography}%

\end{document}